\documentclass[runningheads]{llncs}
\usepackage[T1]{fontenc}
\usepackage{graphicx}
\usepackage{amsmath}
\usepackage{siunitx}
\usepackage{float}
\usepackage[utf8]{inputenc}
\usepackage[english]{babel}
\usepackage{enumitem}
\usepackage{booktabs}
\usepackage{esvect}
\usepackage{graphicx}
\usepackage{microtype}

\title{Effective Stimulus Propagation in Neural Circuits: Driver Node Selection}

\author{Bulat Batuev\inst{1}\orcidID{0000-0002-9322-2635} \and
Arsenii Onuchin\inst{2,3}\orcidID{0000-0002-7811-5831} \and
Sergey Sukhov\inst{4}\orcidID{0000-0002-8966-6030}}

\authorrunning{Batuev et al.}

\institute{Kotelnikov Institute of Radioengineering and Electronics of Russian Academy of Sciences, Moscow, Russia \\
\and
Skolkovo Institute of Science and Technology, Moscow, Russia \\
\and
Laboratory of Complex Networks, Center for Neurophysics and Neuromorphic Technologies, Moscow, Russia\\
\and
Ulyanovsk Branch of Kotelnikov Institute of Radio Engineering and Electronics of Russian Academy of Sciences, Ulyanovsk, Russia \\
\email{buligarmouth@gmail.com}}

\begin{document}

\maketitle

\begin{abstract}
Precise control of signal propagation in modular neural networks represents a fundamental challenge in computational neuroscience. We establish a framework for identifying optimal control nodes that maximize stimulus transmission between weakly coupled neural populations. Using spiking stochastic block model networks, we systematically compare driver node selection strategies—including random sampling and topology-based centrality measures (degree, betweenness, closeness, eigenvector, harmonic, and percolation centrality)—to determine minimal control inputs for achieving inter-population synchronization. 
    
Targeted stimulation of just 10-20\% of the most central neurons in the source population significantly enhances spiking propagation fidelity compared to random selection. This approach yields a $\sim$64-fold increase in signal transfer efficiency at critical inter-module connection densities. These findings establish a theoretical foundation for precision neuromodulation in biological neural systems and neurotechnology applications.

\end{abstract}

\section{Introduction}
Neural synchronization between distinct brain regions underlies critical cognitive processes, including perception, attention, and memory formation \cite{fell2011role},\cite{ward2003synchronous}. Disruptions in inter-regional synchronization are implicated in neurological disorders such as epilepsy, Alzheimer's disease, and schizophrenia \cite{uhlhaas2006neural}. Despite its significance, the structural determinants of synchronization propagation, particularly the role of nodal centrality and inter-areal connectivity, remain incompletely characterized.

Identification of minimal driver inputs required to achieve synchronization offers dual benefits: advancing fundamental understanding of brain network principles and developing targeted therapies for synchronization pathologies. We address this challenge through stochastic block model (SBM) networks, building upon prior work in neural network control \cite{bayati2015self}. Our approach evaluates multiple centrality metrics to determine topological predictors of effective driver nodes.

This research extends beyond previous studies in three key aspects: (1) systematic comparison of six centrality measures for control node identification, (2) explicit quantification of propagation efficiency in spiking neural networks, and (3) analysis of inter-cluster connectivity effects on synchronization transfer. We hypothesize that driver node selection based on specific centrality metrics and inter-cluster connectivity significantly enhances synchronization efficiency in SBM neural networks.

\section{Methods}
\subsection{Network Architecture and Topology Modification}
Neural populations were modeled as undirected stochastic block model (SBM) networks \cite{holland1983stochastic} with two equally sized clusters ($N = 500$ is the total number of neurons, $|V_t| = |V_o| = 250$, where $|V_t|$ and $ |V_o|$ are the sizes of the stimulated $V_t$ and measured $V_o$ populations of neurons). The adjacency matrix $A$ was constructed with intra-cluster connection probability $p_{\mathrm{intra}} = 0.15$ and inter-cluster probability $p_{\mathrm{inter}}$ systematically varied from 0.01 to 0.1. The signal was applied to neurons within the top 10\%, 15\%, or 20\% centrality percentiles ($p_{\mathrm{input}}$) or to 10–20\% of randomly selected neurons within the cluster $V_t$. To overcome SBM's inherent homogeneity and in order to introduce topological diversity, we implemented a centrality boosting procedure defined by:

\begin{equation}
\Delta k(u) = \left\lceil (\beta - 1) \frac{c(u)}{\max_{v \in S} c(v)} k_{\mathrm{ext}}(u) \right\rceil,
\end{equation}
where:
\begin{itemize}[leftmargin=*,noitemsep]
\item $\beta = 1.5$ (50\% centrality boost);
\item $c(u)$ is centrality value of the $u$ vertex;
\item $S$ is the set of selected driver neurons;
\item $k_{\mathrm{ext}}(u)$ is the existing inter-cluster degree.

\end{itemize}

Network density was preserved via compensatory edge removal using the constraint:

\begin{equation}
\sum_{u \in V_t} k_{\mathrm{ext}}(u) = \lfloor p_{\mathrm{inter}} \cdot |V_t| \cdot |V_o| \rfloor .
\end{equation}

Four distinct control node selection strategies were implemented:

\begin{enumerate}[leftmargin=*,label=\textbf{\Roman*.}]
\item \textbf{Top Centrality:} Select $n_{\mathrm{boost}}$ neurons in $V_t$ with maximal $c(u)$
\item \textbf{Proxy Stimulation:} Random selection of neurons in $V_t$ from neighbors of top central nodes in $V_o$
\item \textbf{Random Baseline:} Uniform random selection from $V_t$
\end{enumerate}

\subsection{Centrality Measure Computation}
We evaluated six topological centrality measures that quantify distinct aspects of node influence in networks. For graph $G=(V,E)$ with adjacency matrix $A$ ($A_{uv} = 1$ if edge exists, 0 otherwise), $n = |V|$ nodes, and shortest path distance $d(u,v)$ we define the following centrality measures:

\subsubsection*{Degree Centrality}
Quantifies direct connectivity influence:
\begin{equation}
C_d(u) = \sum_{v \in V} A_{uv}.
\end{equation}
It represents the number of immediate neighbors, indicating local influence capacity.

\subsubsection*{Betweenness Centrality}
Measures brokerage potential in information flow:
\begin{equation}
C_b(u) = \sum_{\substack{s \neq u \neq t \\ s,t \in V}} \frac{\sigma_{st}(u)}{\sigma_{st}},
\end{equation}
where $\sigma_{st}$ is the total shortest paths between $s$ and $t$, and $\sigma_{st}(u)$ is the paths traversing $u$. This centrality identifies nodes functioning as critical bridges.

\subsubsection*{Closeness Centrality}
Characterizes global integration efficiency:
\begin{equation}
C_c(u) = \frac{n - 1}{\sum_{v \neq u} d(u,v)}. \quad \text{(undefined for disconnected components)}
\end{equation}
Nodes with minimal average path distance to others can rapidly disseminate information.

\subsubsection*{Eigenvector Centrality}
Quantifies influence through connectedness to influential neighbors:
\begin{equation}
\mathbf{x} = \lambda_{\max}^{-1} A\mathbf{x}.
\end{equation}
This equation is solved via power iteration to convergence ($\Delta x < 10^{-6}$), where $\lambda_{\max}$ is the largest eigenvalue of $A$. This captures recursive importance propagation.

\subsubsection*{Percolation Centrality}
Models dynamic spreading importance:
\begin{equation}
C_p(u) = \frac{1}{n-2} \sum_{\substack{s \neq u \neq r \\ s,r \in V}} \frac{\sigma_{sr}(u)}{\sigma_{sr}} \frac{x_s^{(t)}}{\sum_{i \in V} x_i^{(t)} - x_u^{(t)}},
\end{equation}
where $x_i^{(t)}$ is the activation probability at time $t$ (simulated over $t=1$ to $10$). This extends betweenness centrality to time-varying processes.

\subsubsection*{Harmonic Centrality}
Robust closeness variant for disconnected graphs:
\begin{equation}
C_h(u) = \frac{1}{n-1} \sum_{v \neq u} \frac{1}{d(u,v)} \quad \left(\frac{1}{d(u,v)} := 0 \text{ if } d(u,v) = \infty\right).
\end{equation}
This centrality penalizes unreachable nodes less severely than standard closeness.

All measures were min-max normalized to $[0,1]$ before analysis. Computations used NetworkX 3.0 \cite{hagberg2008exploring} with Brandes' algorithm for path-based metrics \cite{brandes2001faster}.

\subsection{Neural Dynamics Model}

We consider spiking neural network consisting of leaky integrate-and-fire (LIF) neurons \cite{abbott1999lapicque}. As was mentioned in the previous section, the whole network was divided into two equal populations. The connections between individual neurons within and between populations were established according to the desired adjacency matrix. Each of the two populations contained $80\%$ excitatory and $20\%$ inhibitory neurons. 

The equations for the membrane potential of a single neuron were written as follows:
\begin{equation}
\tau \frac{dv}{dt}
= v_{\mathrm{rest}} - v + R\,I_{\mathrm{ext}} + v_{n} + v_{e} + v_{i},
\end{equation}
where $v_{\mathrm{rest}}$ is the steady-state value of the membrane potential, $R$ is the equivalent membrane resistance, $\tau$ is the characteristic time constant, and $v_{e}$ and $v_{i}$ are changes in the membrane potential caused by synaptic currents from contacting excitatory ($e$) and inhibitory ($i$) neurons:
\begin{align}
v_{e} &\mathrel{+}= \sum_{e} w_{e}\,\delta\bigl(t - t_{e}\bigr),\\
v_{i} &\mathrel{-}= \sum_{i} w_{i}\,\delta\bigl(t - t_{i}\bigr).
\end{align}
To simulate background neural activity, all neurons received independent stimuli $v_{n}$ with a Poisson distribution. Background activity was modeled as Poisson input with rate $\lambda_{\mathrm{bg}} = \SI{1}{\hertz}$ per neuron. Each spike from the Poisson generator causes an instantaneous increase in membrane potential, which can be represented as the sum of delta functions:
\begin{equation}
v_{n} \mathrel{+}= \sum_{n} w_{n}\,\delta\bigl(t - t_{n}\bigr),
\end{equation}
where $t_{n}$ is the time of arrival of the spikes. 
We set
\[
w_{e} = w_{i} = w_{n} = w.
\]
In addition to the noise from external influences, the driver neurons received an additional periodic signal:
\begin{equation}
I_{\mathrm{ext}} = I_{0}\,\sin\bigl(2\pi f t + \phi\bigr).
\end{equation}
In our setting, driver neurons were chosen only amongst the {\em excitatory} neurons of population $V_t$. 

The neuron generated a spike when $v(t)$ reached the excitation threshold $v_{\mathrm{th}}$, after which the potential was instantly reset to $v_{\mathrm{reset}}$. Formally, this is described by the condition
\begin{equation*}
v(t) \ge v_{\mathrm{th}}
\quad\longrightarrow\quad
v(t) = v_{\mathrm{reset}}.
\end{equation*}

All the parameter values used in the simulations are shown in Table 1.

\begin{table}[h]
\centering
\caption{Neuron dynamics simulation parameters}
\begin{tabular}{lcc}
\toprule
Parameter & Symbol & Value \\
\midrule
Membrane capacitance & $C_m$ & \SI{250}{\pico\farad} \\
Membrane resistance & $R_m$ & \SI{80}{\mega\ohm} \\
Resting potential & $v_{rest}$ & \SI{-65}{\milli\volt} \\
Threshold & $v_{th}$ & \SI{-50}{\milli\volt} \\
Reset potential & $v_{reset}$ & \SI{-70}{\milli\volt} \\
Membrane time constant & $\tau_m$ & \SI{20}{\milli\second} \\
Excit. and inhib. weights & $w$ & \SI{1}{} \\
Input current & $I_0$ & \SI{1}{\pico\ampere} \\
frequency & $f$ & \SI{10}{\hertz} \\
phase & $\phi$ & \SI{0}{} \\ 
\bottomrule
\end{tabular}
\end{table}

\subsection{Numerical Experiments}
We conducted several experiments in which we chose specific driver nodes in the population $V_t$ (with a fraction $p_{input}$ of the total amount) and measured neural activity in the population $V_o$.
Several experimental protocols were implemented.
For the first set of experiments, the SBM network was generated with the specified probabilities $p_{\mathrm{intra}}$, $p_{\mathrm{inter}}$.  Next, the centrality measures were computed for the nodes of the $V_t$ population. 10-to-20\% of the nodes with the highest centralities in the population $V_t$ were chosen as driver nodes. The dynamics of the resulting network was simulated for 5 seconds of biological time in Brian~2 neural simulator \cite{stimberg2019brian}. The initial 100 ms warm-up period was excluded from the analysis.

In an alternative setting, the centrality measures were computed for the nodes of the $V_o$ population. The driver nodes were randomly chosen among the nodes of $V_t$ that are neighbors of the top central nodes in $V_o$.

\begin{figure}[h!]
\centering
\includegraphics[width=1\textwidth]{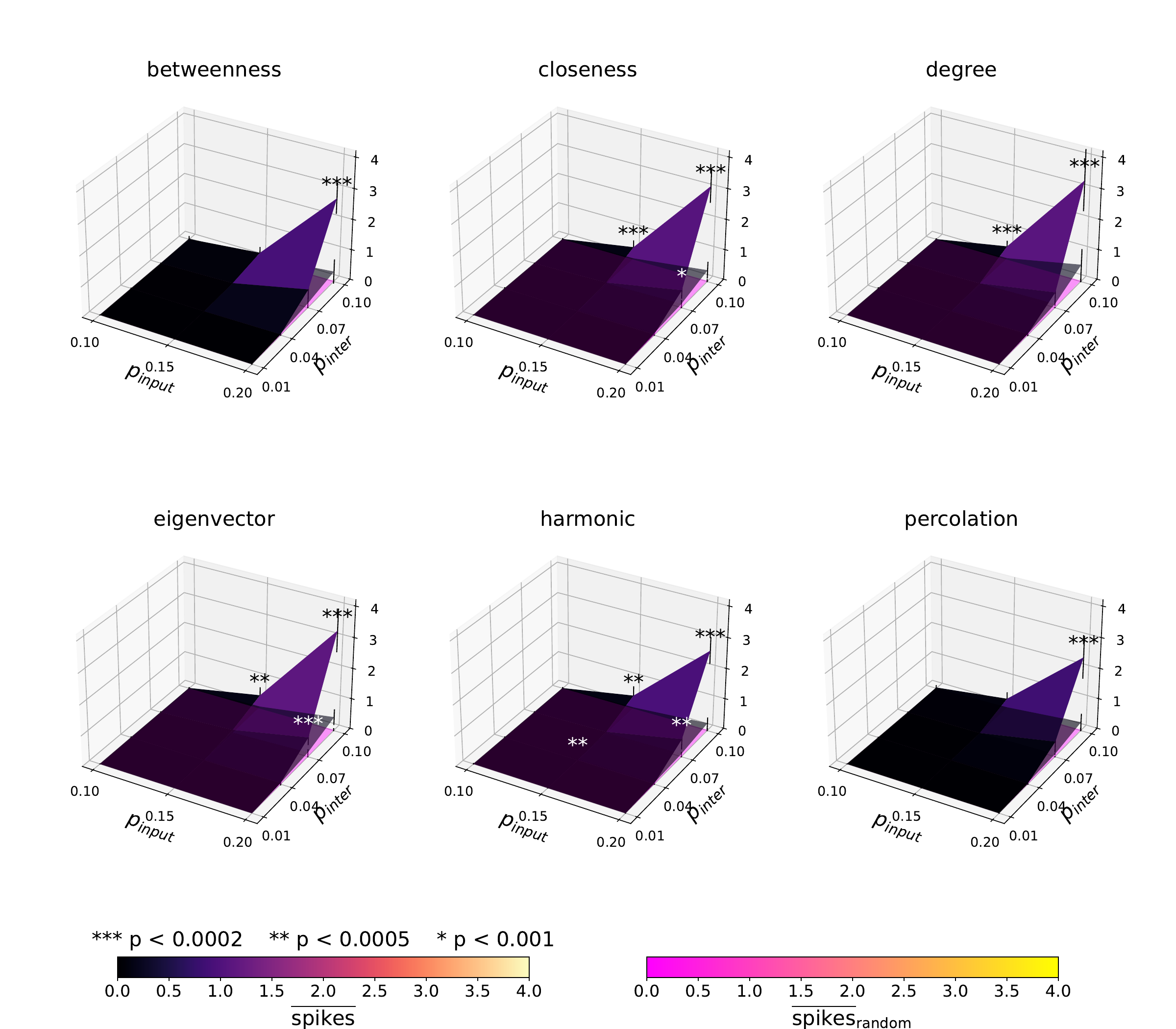}
\caption{Propagation efficiency in non-boosted networks. This figure depicts the distribution of average activity in the $V_o$ population following targeted stimulation of high-centrality nodes within the $V_t$ population. 
Six subgraphs are presented (left to right, top to bottom), corresponding to the following centrality measures: betweenness, closeness, degree, eigenvector, harmonic, and percolation. Each subgraph displays three surfaces representing distinct stimulation conditions: the average activity response in the $V_o$ cluster when stimulating the top-centrality nodes within the $V_t$ cluster; the response when stimulating proxy nodes in the $V_t$ cluster; and a random baseline response from stimulating randomly selected driver node subsets ($p_{input} = 10\%, 15\%, 20\%$) in the $V_t$ cluster. The top-centrality and proxy stimulation results share a single colormap scale but are visually distinguished by differential transparency, while the random baseline employs a separate colormap scale as visually indicated}
\label{fig:nonboosted_perf}
\end{figure}

The above experimental settings were compared to the baseline experiment in which the driver nodes were chosen in the population $V_t$ randomly. The simulations were repeated 20 times for each set of parameters. The results of simulations are shown in Fig.~\ref{fig:nonboosted_perf}.

\begin{figure}[h!]
\centering
\includegraphics[width=1\textwidth]{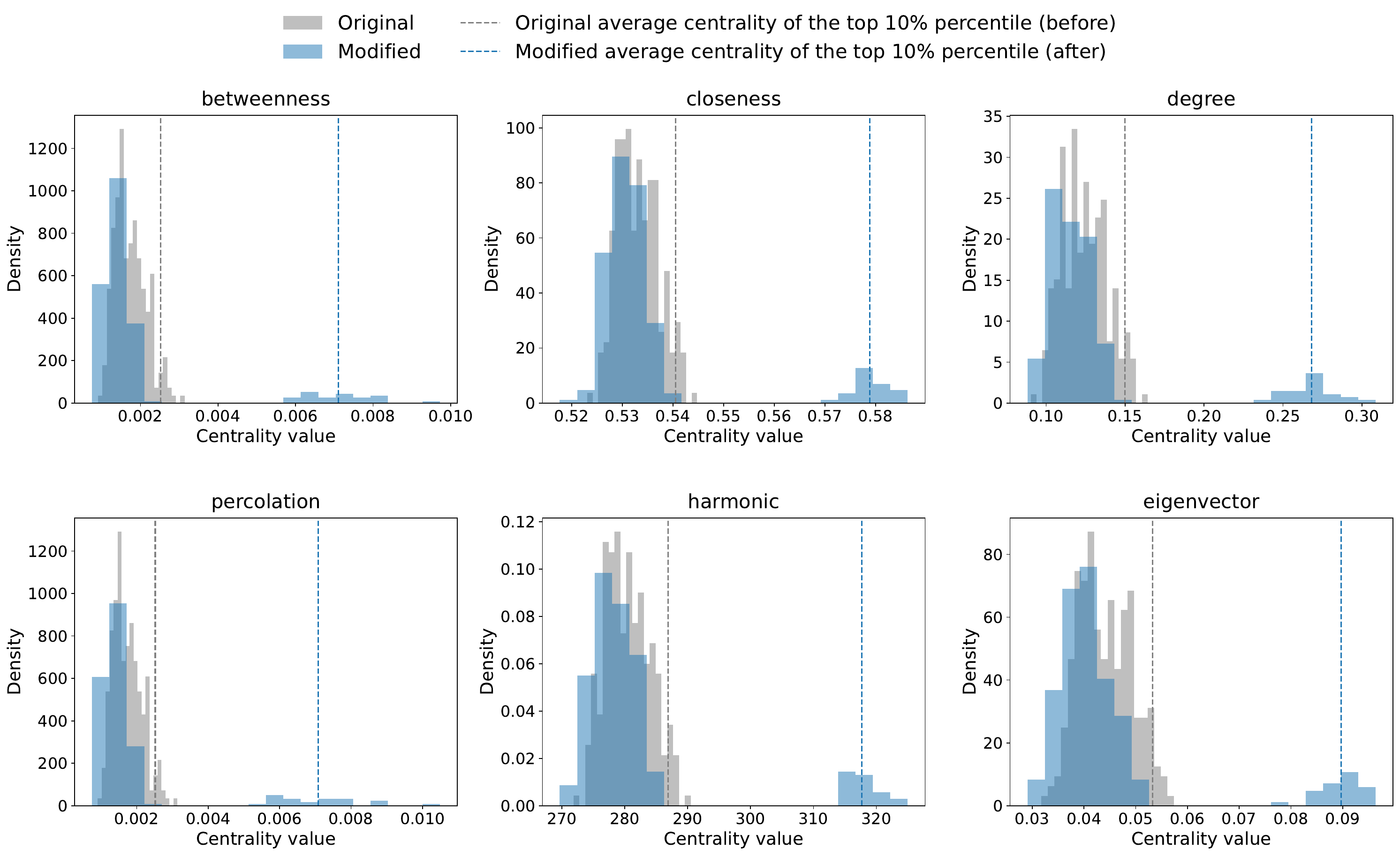}
\caption{The distribution of centrality metrics values for network nodes is shown before boosting (Original) and after boosting (Modified). In the boosted case, a distinct tail emerges representing the nodes with significantly increased centrality of the selected type}
\label{fig:centrality_destr}
\end{figure}

\begin{figure}[h!]
\centering
\includegraphics[width=0.9\textwidth]{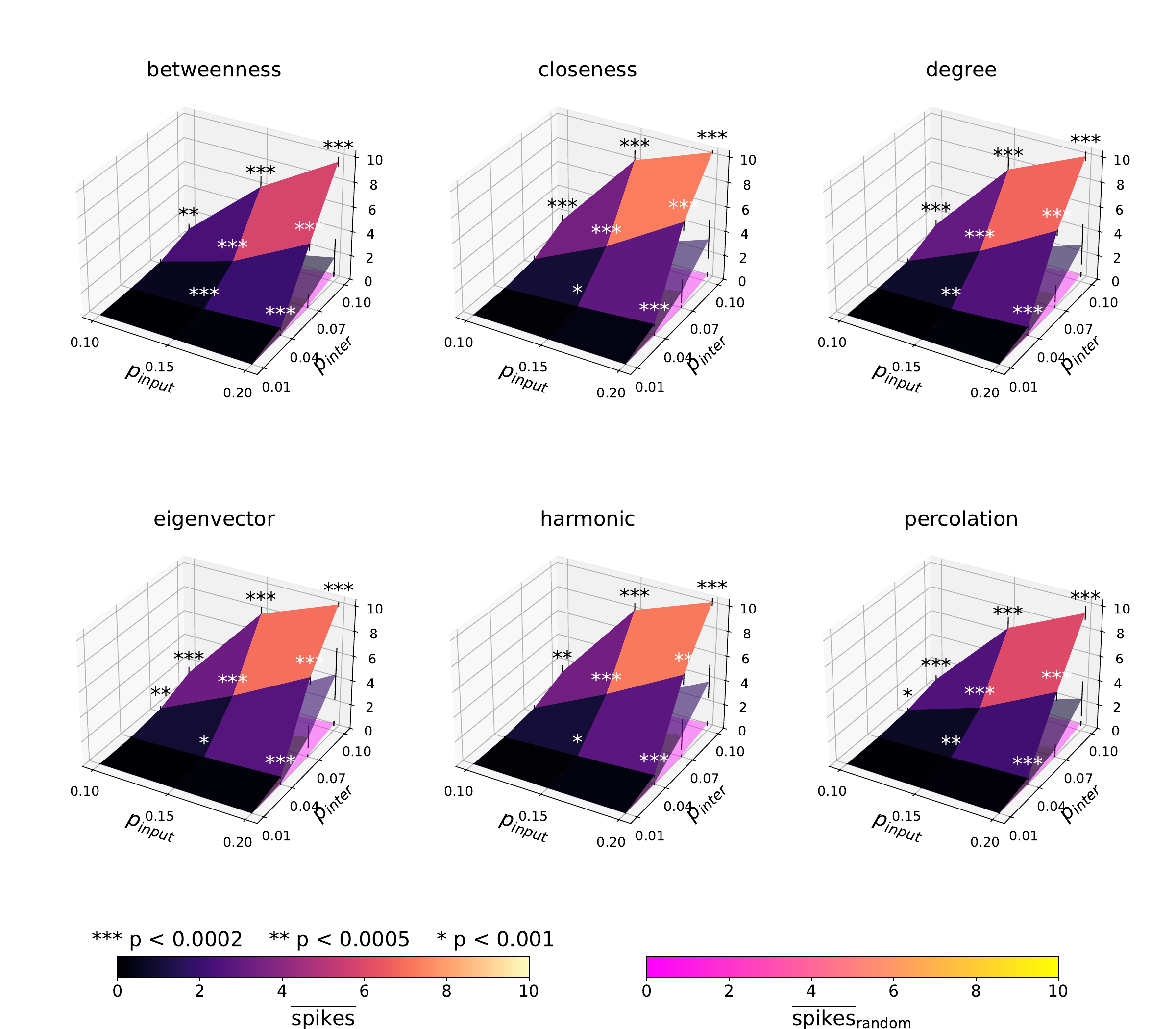}
\caption{Propagation efficiency in boosted networks. This figure depicts the distribution of average activity in the $V_o$ population following targeted stimulation of high-centrality nodes within the $V_t$ population.
Six subgraphs are presented (left to right, top to bottom), corresponding to the following centrality measures: betweenness, closeness, degree, eigenvector, harmonic, and percolation. Each subgraph displays three surfaces representing distinct stimulation conditions: the average activity response in the $V_o$ cluster when stimulating the top-centrality nodes within the $V_t$ cluster; the response when stimulating proxy nodes in the $V_t$ cluster; and a random baseline response from stimulating randomly selected node subsets ($p_{input} = 10\%, 15\%, 20\%$) in the $V_t$ cluster. The top-centrality and proxy stimulation results share a single colormap scale but are visually distinguished by differential transparency, while the random baseline employs a separate colormap scale as visually indicated}
\label{fig:boosted_perf}
\end{figure}

In the second set of experiments, the experimental procedure was similar to the previous one except the additional boosting applied to the $n_{boost}$ nodes with the highest centrality (Fig.~\ref{fig:centrality_destr}):
\begin{itemize}
\item Topology of the network was modified by adding $\Delta k$ edges.
\item The increased density of edges was compensated by random edge removal.
\end{itemize}
The results of these simulations are shown in Fig.~\ref{fig:boosted_perf}.

\section{Results}


The control strategies described in the previous section demonstrated different efficiencies in activity propagation between populations.
In the absence of centrality boosting, the driver nodes selection strategy significantly impacted the activity propagation efficiency. Crucially, stimulation of the top centrality nodes substantially outperformed proxy stimulation across all parameter set elements (see Fig.~\ref{fig:nonboosted_perf}).

In non-boosted networks, our analysis reveals fundamental constraints in stimulus propagation through structural connectomes. Below the critical percolation threshold ($p_{\mathrm{inter}} < 0.04$), no strategy elicited significant propagation -- indicating a fundamental physical constraint on information transmission in sparse connectomes. Beyond this threshold, strategy-dependent effects emerged:

\begin{enumerate}
    \item \textbf{Medium connectivity regime} ($p_{\mathrm{inter}} = 0.07$):  
    Stimulation of top eigenvector centrality nodes achieved moderate propagation ($0.64 \pm 0.24~\text{Hz}$), consistent with their role in global network integration. Interestingly, proxy stimulation targeting the central nodes in the kinship of the betweenness near the central regions showed severely attenuated efficacy ($0.15 \pm 0.14~\text{Hz}$), representing a 4.26-fold reduction versus optimal targets. This performance approached the random baseline ($0.01 \pm 0.01~\text{Hz}$), demonstrating that anatomical adjacency to hubs is insufficient for effective control.
    
    \item \textbf{High connectivity regime} ($p_{\mathrm{inter}} = 0.10$):  
    Degree-centrality targeting maximized propagation ($3.33 \pm 0.98~\text{Hz}$), leveraging local integration capacity. Proxy stimulation using degree central nodes near the hubs remained markedly inefficient ($0.57 \pm 0.51~\text{Hz}$, a 5.84-fold reduction). The persistent strategy gap across densities suggests that conventional hub-centric approaches neglect critical control mechanisms beyond local connectivity.
\end{enumerate}

\textbf{Topological Enhancement Reshapes Control Landscapes}
Centrality boosting amplified global propagation efficiency and altered network dynamics:

    

\begin{enumerate}
    \item \textbf{Threshold modulation}:  
    Boosted networks exhibited significant propagation amplification. Compared to non-boosted case, there is 3.15$\times$ increase in propagation efficiency compared to non-boosted networks at $p_{\mathrm{inter}}=0.10$, with the mean firing rate in $V_o$ rising from $3.33 \pm 0.98$ Hz to $10.5 \pm 0.13$ Hz under optimal driver selection. This demonstrates that targeted neuroplasticity can reconfigure information routing pathways to overcome physical connectivity constraints.
    
    \item \textbf{Convergence of control strategies}:
    \begin{itemize}
        \item At $p_{\mathrm{inter}} = 0.07$, closeness-centrality targeting dominated ($7 \pm 0.69~\text{Hz}$), while proxy stimulation remained ineffective ($1.8 \pm 1.23~\text{Hz}$, 3.89-fold reduction)  
        \item At $p_{\mathrm{inter}} = 0.10$, closeness-centrality again prevailed ($10.5 \pm 0.13~\text{Hz}$), but proxy stimulation showed relative improvement ($4.7 \pm 2.07~\text{Hz}$, 2.23-fold reduction)
    \end{itemize}
\end{enumerate}

\textbf{Interpretive Synthesis}
The persistent >55\% efficacy gap in proxy stimulation -- even in boosted networks -- underscores that spatial proximity to hubs cannot compensate for the absence of \textit{topological position}. Optimal targets consistently exhibited high closeness centrality, reflecting their capacity to integrate information through short functional paths. By contrast, the modest improvement in proxy efficacy under high connectivity with boosting suggests that enhanced network integration partially compensates for suboptimal target selection. 

\begin{figure}[h!]
\centering
\includegraphics[width=0.95\textwidth]{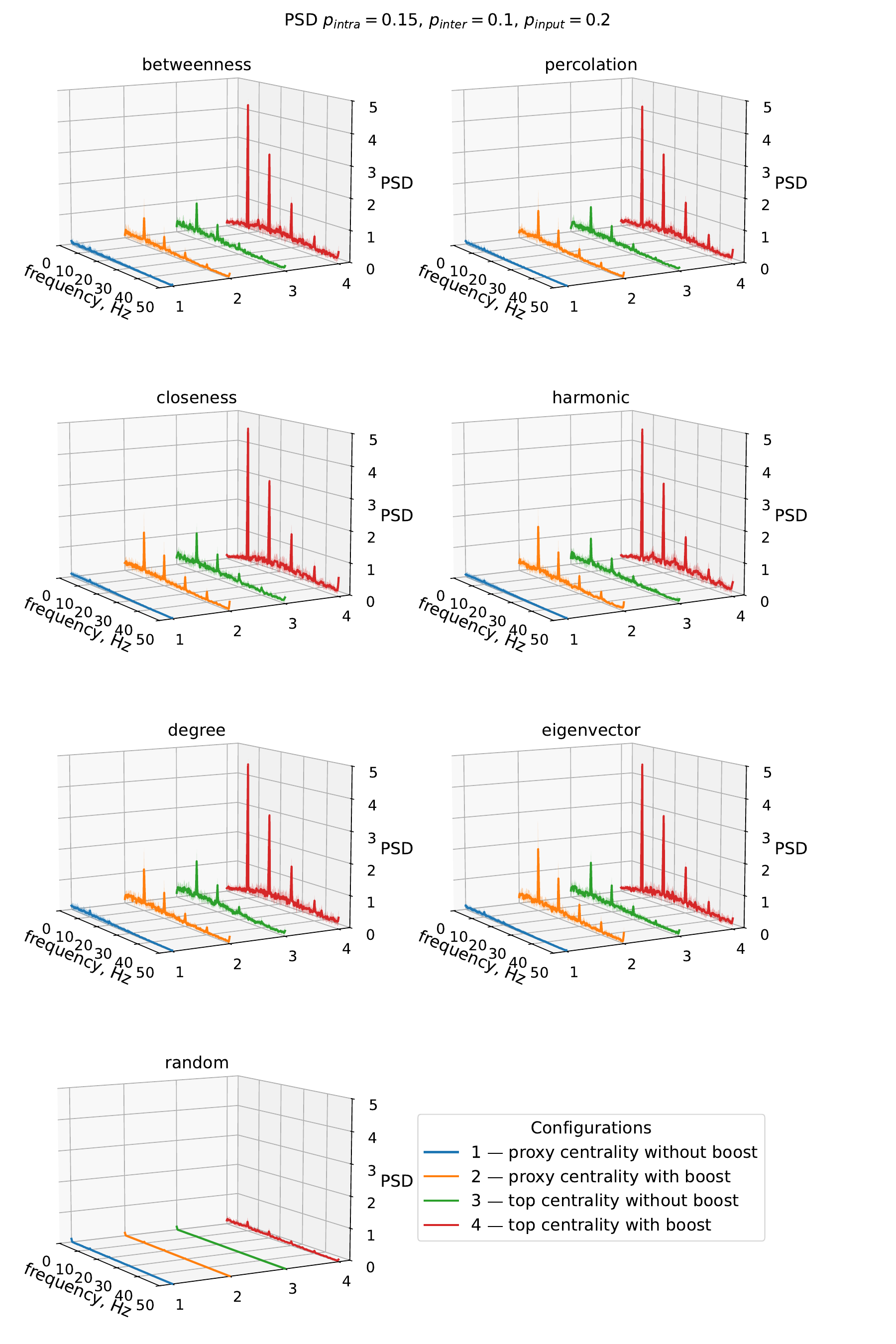}
\caption{Dynamic propagation characteristics. (A) Non-boosted networks: Top-centrality shows narrowband enhancement at driving frequency (10 Hz). (B) Boosted networks: All strategies exhibit spectral peaks but with varying signal-to-noise ratios (SNR: top=8.27 dB, proxy=7.49 dB)}
\label{fig:psd}
\end{figure}

\section{Discussion}
Our results demonstrate that topological centrality provides an effective criterion for driver nodes selection in modular neural networks. The superior performance of top-centrality nodes stems from their dual advantage: (1) high integration within source populations enables efficient input amplification, and (2) preferential connection to inter-module bridges facilitates propagation.

The limited efficacy of proxy stimulation reveals that spatial proximity to target hubs is insufficient for effective control, aligning with recent findings that optimal control requires both local integration and global bridging capabilities \cite{gu2015controllability}. Spectral analysis confirms that centrality-based selection preserves the temporal fidelity of transmitted signals (Fig.~\ref{fig:psd}).

Notably, the performance gap between selection strategies was most pronounced at critical $p_{\mathrm{inter}}$ values (0.07), corresponding to theoretical percolation thresholds for modular networks. This indicates topological control strategies are particularly valuable near network phase transitions, precisely where biological neural systems typically operate.

\section{Conclusion}
We establish that targeted stimulation of top-centrality neurons significantly outperforms both proximity-based and random selection strategies for activity propagation between neural populations. Key findings demonstrate:

\begin{itemize}
    \item \textbf{Critical operating regime}: Centrality-based selection achieves maximum advantage ($64\times$ efficiency gain over random baseline, $p_{\mathrm{value}} < 10^{-9}$) near network percolation thresholds;
    
    \item \textbf{Metric-specific performance}: Closeness and harmonic centrality consistently outperformed other metrics, with closeness centrality demonstrating particular robustness under topological enhancement ($>5\times$ advantage over proxy strategies);
    
    \item \textbf{Mechanistic distinction}: Centrality-based stimulation preserves temporal signal fidelity while proxy strategies amplify broadband noise.
\end{itemize}

These results establish that effective neural control requires \textit{global topological influence} rather than mere proximity to target hubs. The identified performance peak at critical connection densities provides a principled framework for precision neuromodulation in neurological disorders characterized by synchronization deficits. Future work should extend this approach to dynamic adaptive networks and validate findings in \textit{in vivo} recordings.

\subsubsection*{Acknowledgments}
The study was supported by the Russian Science Foundation, project number 24-21-00470.

\bibliographystyle{splncs03_unsrt}
\bibliography{references}    
\end{document}